\documentclass[lettersize,journal]{IEEEtran}

\usepackage{amsmath,amsfonts}
\usepackage[bookmarks,unicode,hidelinks]{hyperref}
\usepackage{forest}
\usepackage{tikz}
\usepackage{caption}
\usepackage{subcaption}
\usepackage[strings]{underscore}
\usepackage{cite}
\usepackage[final]{microtype}
\usepackage{orcidlink}
\usepackage[nospread]{flushend}

\usepackage[disable]{todonotes}
\usepackage{enumitem}

\newtheorem{definition}{Definition}

\newcommand*\halfcirc[1][black]{
    \begin{tikzpicture}
        \draw[fill=#1, draw=#1] (0,0)-- (90:1ex) arc (90:270:1ex) -- cycle ;
        \draw[draw=#1] (0,0) circle (1ex);
\end{tikzpicture}}
\newcommand*\fullcirc[1][black]{\tikz\draw[fill=#1, draw=#1] (0,0) circle (1ex);} 

\definecolor{color1}{HTML}{189799}
\definecolor{color2}{HTML}{39E615}
\definecolor{color3}{HTML}{E6C211}
\definecolor{color4}{HTML}{9E0C9E}
\definecolor{color5}{HTML}{E6540F}
\definecolor{color6}{HTML}{0F109E}

\newcommand\fullallcircle[0]{\fullcirc[color1]}
\newcommand\halfallcircle[0]{\halfcirc[color1]}
\newcommand\fullenvcircle[0]{\fullcirc[color2]}
\newcommand\halfenvcircle[0]{\halfcirc[color2]}
\newcommand\fullstaticcircle[0]{\fullcirc[color3]}
\newcommand\halfstaticcircle[0]{\halfcirc[color3]}
\newcommand\fullpedestriancircle[0]{\fullcirc[color4]}
\newcommand\halfpedestriancircle[0]{\halfcirc[color4]}
\newcommand\fulldynamiccircle[0]{\fullcirc[color5]}
\newcommand\halfdynamiccircle[0]{\halfcirc[color5]}
\newcommand\fulldrivingcircle[0]{\fullcirc[color6]}
\newcommand\halfdrivingcircle[0]{\halfcirc[color6]}

\newcommand\code[1]{\texttt{#1}}
\newcommand\bind[2]{\ensuremath{\downarrow^{#1}_{#2}\hspace{-.3em}}}

\newcommand{\odd}{\ensuremath{\mathbb{T}}}
\newcommand{\tscclasses}{\ensuremath{\mathcal{T}}}

\newcommand{\sig}{\sigma}
\newcommand{\rels}{\mathcal{R}}
\newcommand{\arity}{\mathrm{ar}}

\newcommand{\domain}{\ensuremath{\mathcal{D}}}
\newcommand{\struct}{\ensuremath{\mathfrak{D}}}
\newcommand{\structseq}{\ensuremath{\vec{\struct}}}
\newcommand{\timestampseq}{\ensuremath{\vec{\tau}}}

\newcommand{\scenario}{\ensuremath{\overline{\mathcal{S}}}}
\newcommand{\scenarios}{\ensuremath{\mathfrak{S}}}

\newcommand{\grp}[1]{\left( #1 \right)}

\newcommand{\set}[1]{\lbrace #1 \rbrace}
\newcommand{\tuple}[1]{\langle #1 \rangle}

\begin{document}

\title{Tree-Based Scenario Classification: A Formal Framework for Coverage Analysis on Test Drives of Autonomous Vehicles}

\author{
	\IEEEauthorblockN{Till Schallau\IEEEauthorrefmark{1} \orcidlink{0000-0002-1769-3486}, Stefan Naujokat\IEEEauthorrefmark{1} \orcidlink{0000-0002-6265-6641}, Fiona Kullmann\IEEEauthorrefmark{1} \orcidlink{0000-0001-5858-0659}, and Falk Howar\IEEEauthorrefmark{1}\IEEEauthorrefmark{2} \orcidlink{0000-0002-9524-4459}}\\
	\IEEEauthorblockA{\IEEEauthorrefmark{1}TU Dortmund University, Dortmund, Germany}\\
	\IEEEauthorblockA{\IEEEauthorrefmark{2}Fraunhofer ISST, Dortmund, Germany}\\
	\{till.schallau, stefan.naujokat,  fiona.kullmann, falk.howar\}@tu-dortmund.de
}

\maketitle

\begin{abstract}
	Scenario-based testing is envisioned as a key approach
	for the safety assurance of autonomous vehicles. 
	In scenario-based testing, relevant (driving) scenarios
	are the basis of tests. Many recent works focus on 
	specification, variation, generation and execution of individual 
	scenarios. In this work, we address the open challenges 
	of classifying sets of scenarios and measuring coverage of theses scenarios in recorded test drives.
	Technically, we define logic-based classifiers that 
	compute features of scenarios on complex data streams 
	and combine these classifiers into feature trees that 
	describe sets of scenarios. 
	We demonstrate the expressiveness and effectiveness of our approach 
	by defining a scenario classifier for urban driving and evaluating it
	on data recorded from simulations.
\end{abstract}


\section{Introduction}
\label{sec:intro}

One of the open challenges in the development of autonomous driving software is
assuring its safety~\cite{Mariani2018OverviewAutonomousVehicles}. It has long been established that statistical arguments on
the performance of the complete system (e.g., caused fatalities per million
miles) are not attainable in
practice~\cite{Junietz2018EvaluationDifferentApproaches,Kalra2016DrivingSafetyHow}.
The billions of miles that would have to be driven without failures are simply
not feasible for every new vehicle or software update. For several years now,
the focus of research has been on structured approaches to assuring the safety
of autonomous driving functions instead~\cite{Mauritz2016AssuringSafetyAdvanced, Felbinger2019ComparingTwoSystematic}.

The recently published ISO 21448~\cite{ICS2022RoadVehiclesSafety} norm (Safety
of the Intended Functionality) transfers the conceptual framework of system
safety approaches (e.g., ISO 26262~\cite{ICS2018RoadVehiclesFunctional}) to the
assurance of a vehicle’s safety under all environmental conditions and possible
faults that are triggered by the environment~\cite{Saberi2020SotifBlackSwans}.
Basically, the idea is to identify relevant driving situations and potential
triggers and then use these as a basis for testing the safety of a vehicle or
its driving software. Many recent works focus on defining notions of
safety~\cite{Schuett2022TaxonomyQualitySimulation},
formalizing what constitutes
scenarios~\cite{Schuldt2013EffizienteSystematischeTestgenerierung,
Scholtes20216LayerModel, Ulbrich2015DefiningSubstantiatingTerms,
Nalic2020ScenarioBasedTesting}, and on testing safety in specified
scenarios~\cite{Riedmaier2020SurveyScenarioBased}.

Recent standardization efforts target the specification of so-called
operational design domains (ODDs)~\cite{BSI2020OperationalDesignDomain} that
define the anticipated environmental conditions for autonomous vehicles at a
high level (e.g., weather conditions, road types and parameters, etc.). To
combine works and results on testing individual scenarios into compelling
arguments about the safety of an autonomous vehicle in its operational design
domain, we need tools for describing sets of relevant scenarios in some ODD and
methods for analyzing coverage of these scenarios in driving tests as, e.g.,
stated in the ASAM OpenODD concept~\cite{StandardizationofAutomation2021AsamOpenoddConcept}.

In this paper, we present an approach to the specification of sets of scenarios
through classifiers for features of scenarios in the set. These classifiers can
then be used to identify observed scenarios in recorded test drives. Moreover,
we can compute the set of possible combinations of features from our
specification. This enables us to provide coverage metrics and to identify
counterfactual scenarios, i.e., scenarios that were not observed but could be
observed. Technically, we use logic-based classifiers that identify features of
scenarios on complex data streams and combine these classifiers into feature
trees that describe sets of scenarios, emerging from the combinatorial
combination of features. We extend an existing modal logic to express features
that can be found in ODD standards, in the 6-layer model of driving
scenarios~\cite{Scholtes20216LayerModel, Nalic2020ScenarioBasedTesting}, and in the classification of
driving maneuvers (e.g., intersection, traffic light present, light rain,
oncoming traffic, left-turn maneuver, etc.).

We demonstrate the expressiveness and effectiveness of our approach in a case
study: We specify a small set of features and use test drives in a randomized
simulation to analyze the observed scenario classes and the coverage that can be achieved in
this setup. We also show how coverage can be decomposed and only analyzed for
individual features or layers of the 6-layer model.

\begin{figure*}[t!]
  \begin{subfigure}{0.5\linewidth}
      \centering
      \includegraphics[width=.78\linewidth]{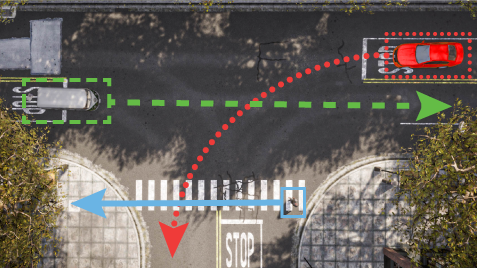}
      \caption{The traffic situation as seen from a birds eye view}
      \label{fig:sfig2}
  \end{subfigure}\hfill
  \begin{subfigure}{0.5\linewidth}
      \centering
      \includegraphics[width=.78\linewidth]{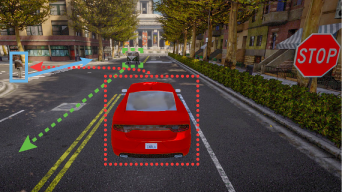}
      \caption{The traffic situation as seen from ego's perspective}
      \label{fig:sfig1}
  \end{subfigure}
  \caption{Screenshots taken in CARLA~\cite{Dosovitskiy2017CarlaOpenUrban} showing a traffic situation in which the
  ego vehicle (red, dotted) stops at a stop sign. The trajectory of the planned
  left turn is crossed by an oncoming vehicle (green, dashed) and by a
  pedestrian (blue, solid) crossing the street.}
  \label{fig:figMotEx}
\end{figure*}

Summarizing, the contribution of this paper is threefold:
\begin{enumerate}
  \item We present a formal logic for describing properties over recorded
  sequences of scenes. 
  The logic extends upon existing temporal logics in
  multiple aspects that are essential for concise specifications that work on
  field-recorded data: firstly, the logic allows fuzzy specifications (in the
  spirit of: “most of the time”); secondly, it is defined over complex
  structured domains for capturing scenes 
  (cf. Sect.~\ref{sec:logic}).

  \item We present a method for classifying sets of scenarios that is
  conceptually inspired by recent works and standardization efforts around
  operational design domains (ODDs) and technically inspired by feature models~\cite{Schobbens2006FeatureDiagramsSurvey},
  where features of scenarios are specified using the presented logic over
  sequences of scenes. To the best of our knowledge, this is one of 
  the first approaches that addresses classification and analyses on sets of 
  different scenarios (cf. Sect.~\ref{subsec:scenario-classifiers}).

  \item The specification of features and formal models for sets of scenarios
  enable several quantitative and qualitative analyses on recorded driving
  data, e.g., scenario coverage, missing scenarios, missing combination of
  features, and distribution of combinations of features. In contrast to other
  works, these metrics focus on sets of scenarios instead of on parameter
  ranges within one scenario (cf. Sect.~\ref{sec:coverage}).
\end{enumerate}
The ultimate goal of this work is to get a
  handle on the task of specifying, selecting, and prioritizing relevant
  scenarios and representative combinations of environmental conditions across
  all scenarios.

\medskip
\noindent
\textbf{Outline.}
The paper is structured as follows. Section~\ref{sec:example} outlines an
example that motivates our approach. The formal logic for defining scenario
properties is introduced in Sect.~\ref{sec:logic}. 
Section~\ref{sec:tscAndMetrics} then introduces the formalism for scenario classifier trees
and the calculation of coverage metrics and analyses
on such trees. The
results of our case study are presented in Sect.~\ref{sec:evaluation}, which
is followed by a discussion on related work in Sect.~\ref{sec:relatedWork}. The paper
concludes in Sect.~\ref{sec:conclusion}.

\medskip
\noindent
\textbf{Reproduction Package.} 
For the experiments conducted i
n our case study, a reproduction package is available on Zenodo~\cite{Schallau2023}.

\section{Motivational Example}
\label{sec:example}    

We illustrate our approach for the task of analyzing test drives in
an urban environment. We assume to have a database of recorded test
drives. Recordings consist of sequences of \emph{scenes} and are 
split into meaningful \emph{segments}, e.g., based on regions of a map.
A single scene is the snapshot of the state and observed 
environment of the ego vehicle, 
comprising map data, position and velocity of the 
ego vehicle, stationary objects, and moving objects around 
the ego vehicle. Segments (i.e., sequences of scenes) are recorded
at fixed (e.g., $0.5$-second) intervals. 

Our task is now to decide if this database contains test drives that
cover enough relevant \emph{scenarios} (i.e., archetypes of
driving situations), or at least to identify and classify the encountered
scenarios. A scenario, in this case, would be a basic driving task,
like making an unprotected left turn on a three-way intersection,
and it could have variants, (e.g., presence of 
oncoming traffic or pedestrians). 

Figure~\ref{fig:figMotEx} shows an example of a scene from a segment 
in which three road users are situated on a T-junction. 
The ego vehicle, which is marked by the red box, is planning
to turn left. It is currently stopped at the stop line
of the stop sign on the ego vehicle's lane.
The car marked with the green box is following
its lane, going straight over the junction and crossing the trajectory 
of the ego vehicle. The destination lane of ego contains a crosswalk
on which a pedestrian, marked in solid blue, is currently crossing the road.
The pedestrian is also crossing the trajectory of the ego vehicle.

When analyzing the segment, specific maneuvers, 
environmental properties, and features can be observed from the 
viewpoint of the ego vehicle:
road type is \emph{T-junction}; ego is \emph{turning left};
there is \emph{oncoming traffic}; a \emph{stop sign} is present;
the ego vehicle does \emph{stop at the stop line},
since it \emph{must yield} to another vehicle; 
a \emph{pedestrian is crossing} the destination
lane; the weather is \emph{sunny} during \emph{daytime}.
These features can be formally described by formulas
in a temporal first-order logic over sequences of scenes. 
A set of features can then be used to classify segments:
the combination of features that hold defines the scenario class.
The segment then is one concrete instance of this scenario class.

Assuming that features are not entailed by other features, 
we generate $2^n$ scenario classes with $n$ features. 
For the more realistic case that some dependencies exist between 
features (e.g., no overtaking without multiple lanes), 
we can use trees to model taxonomies of features and still 
compute possible scenario classes and check if they exist in 
our data.
Possible variations of features in the example could 
be: \emph{the ego vehicle drives straight instead of turning left}, 
\emph{there is no oncoming traffic}, or \emph{no pedestrian is crossing the road}. 
For the sake of simplicity, we neglect the other properties for the following calculation.
Based on these three variations,
a total of ${2^3=8}$ possible scenario classes are observable.
We can use this information to compute missed scenario classes or to 
measure scenario class coverage for our database of test drives.
In our example, one scenario class was observed.
Given the eight possible scenario classes, we obtain
a scenario class coverage of $12.5\%$. 
The following two sections formalize these concepts.

\section{A Temporal Logic for Properties of Scenarios}
\label{sec:logic}

We base our classifiers for scenarios on the environment representation 
that is usually produced by the perception sub-system of an autonomous
vehicle: a map of the road network and typed objects with positions, 
velocities, and observed states.
To express properties of recorded sequences of scenes (i.e., momentary
snapshots of the environment of the ego vehicle), we need a formal logic
that can express properties in individual scenes as well properties between
objects in multiple different scenes. Examples are distances between objects in
one scene, or the fraction of scenes in a sequence in which a leading 
vehicle is present. We introduce such a logic and then use it 
for defining classifier trees that  express sets of scenarios in terms 
of features in the scenarios (cf. Sect.~\ref{subsec:scenario-classifiers}).

We use logic structures to describe scenes over a given 
signature of domain-specific functions and relations
(e.g., positions, lanes, vehicles, velocities, etc.).
We introduce
\emph{CMFTBL} (Counting Metric First-Order Temporal Binding Logic),
a metric first-order logic for modeling time that
extends \emph{MFOTL} (Metric First-Order Temporal 
Logic)~\cite{Mueller2009TheoryApplicationsRuntime, Basin2015MonitoringMetricFirst},
while focusing on finite traces of states.
In particular, we extend MFOTL by a
\emph{minimum prevalence operator} that allows us to express that a property (or
sub-formula) holds for a certain fraction of all future states (within the finite trace).
We also introduce a \emph{binding operator}
that stores an evaluation of a term into a variable, so that the 
result of this evaluation can
be accessed in operator contexts of future states.  
While the former extends the expressiveness of MFOTL,
the second one is a shorthand for existentially 
quantified formulas of a certain form.

A signature $\sig$ is a tuple 
$\tuple{\mathcal{C}, \mathcal{F}, \mathcal{R}, \arity}$,
where $\mathcal{C}$ is a set of named constants,
$\mathcal{F}$ is a set of function symbols, 
$\mathcal{R}$ is a set of relation symbols, and 
$\arity: (\mathcal{F}\cup\mathcal{R}) \mapsto \mathbb{N}_0$ 
defines an arity for each 
function symbol $f \in\mathcal{F}$ and
relation symbol $r \in\mathcal{R}$. 
A $\sig$-structure $\struct$ is a pair 
$\tuple{\domain, I}$ of a domain $\domain$
and interpretations of constants, functions,
and relations with $I(c) \in \domain$
for $c\in \mathcal{C}$,
$\arity(f)$-ary function $I(f):~\domain^{\arity(f)} \to \domain$ 
for $f \in \mathcal{F}$, and 
$I(r) \subseteq \domain^{\arity(r)}$
for $r \in \mathcal{R}$.

An interval of the set of non-empty intervals $\mathcal{I}$ over $\mathbb{N}$ can be written as
$[b,b'):=\{a\in\mathbb{N}|b\leq a<b'\}$, where
$b\in \mathbb{N}, b'\in\mathbb{N}\cup\{\infty\}$ and $b<b'$.

\emph{CMFTBL formulas} over the signature $\sig$, intervals $\mathcal{I}$, and the countably 
infinite set of variables $\mathcal{V}$ (assuming $\mathcal{V} \cap (\mathcal{C} \cup \mathcal{F} \cup \mathcal{R}) = \emptyset$) 
are inductively defined as follows:

\begin{enumerate}[label=(\roman*)]

  \item A \emph{term} $t$ is either a constant $c$, a variable $v$, or for $f\in \mathcal{F}$ and terms $t_1,\cdots,t_{\arity(f)}$ the application $f(t_1,\cdots, t_{\arity(f)})$.

  \item For $r \in \mathcal{R}$ and terms $t_1,\cdots,t_{\arity(r)}$, the predicate\\
  $r(t_1,\cdots, t_{\arity(r)})$ is a \emph{formula}.

  \item For $x \in \mathcal{V}$ and $d \in \domain$, if $t$ is a term, $\varphi$ and $\psi$ are formulas, then $(\lnot \varphi), (\varphi \lor \psi)$, $(\exists x:\varphi)$, and $(\bind{t}{x}\varphi)$ are formulas,
  where $\bind{t}{x}$ \hspace{.2em} evaluates $t$ in the current state and binds the result to variable $x$.

  \item For $I \in \mathcal{I}$ and $p \in \mathbb{R}$, if $\varphi$, and $\psi$ are formulas then 
  next $(\circ_I \varphi)$, 
  until $(\varphi U_I \psi)$, and 
  min. prevalence $(\nabla_I^p \psi)$ are formulas.
\end{enumerate}

While the semantics of \emph{MFOTL} is defined over infinite sequences, we restrict our 
attention and definitions to finite sequences. 
The pair $\tuple{\structseq, \timestampseq}$ is a \emph{finite temporal
structure} over the signature $\sig$, where $\structseq = (\struct_0,
\struct_1, \cdots, \struct_n)$ is a finite sequence of structures (i.e.,
scenes) over $\sig$ and $\timestampseq=(\tau_0, \tau_1, \cdots, \tau_n)$ is a
finite sequence of non-negative rational numbers $\tau_i \in \mathbb{Q}^+$ with length $n$.
The elements in the sequence \timestampseq{} are (increasing) \emph{time
stamps}. Furthermore, the interpretations of relations
$r^{\struct_0}, r^{\struct_1},
\cdots, r^{\struct_n}$ in a temporal structure $\tuple{\structseq,
\timestampseq}$ corresponding to a predicate symbol $r \in \mathcal{R}$ may
change over time. The same is true for functions. 
Constants $c\in \mathcal{C}$ and the domain \domain, on the other hand, do not change over time. More formally, we assume for all $0 \leq i < n$ that 
$\tau_i < \tau_{i+1}$ and
for $\struct_i=\tuple{\domain_i, I_i}$ and $\struct_{i+1}=\tuple{\domain_{i+1}, I_{i+1}}$ that $\domain_i = \domain_{i+1}$.
Moreover, $c^{\struct_i} = c^{\struct_{i+1}}$ for each constant symbol $c\in \mathcal{C}$.  

A \emph{valuation} is a mapping 
$v:\mathcal{V}\rightarrow \domain$ 
from variables to domain elements. We write $v [x \mapsto d]$ for the valuation $v$ that maps $x$ to $d$. All other variables are not affected in the valuation $v$. We abuse notation by applying a valuation $v$ also to constant symbols $c\in \mathcal{C}$, with $v(c)=c^{\struct}$.\\

We evaluate term $t$ for valuation $v$ and structure \struct, denoted by 
$\beta[t,v,\struct]$ as follows. For constants and variables $x$, let
$\beta[x,v,\struct] = v(x)$. For function application $a=f(t_1,\cdots, t_{\arity(f)})$,
let $$
\beta[a, v, \struct] = f^\struct( \beta[t_1,v, \struct],\cdots, \beta[t_{\arity(f)},v, \struct]).$$

We define the semantics of CMFTBL in terms of the relation $(\structseq,
\timestampseq, v, i)\models_{CMFTBL} \varphi$ inductively in
Table~\ref{tab:cmftbl}, where $\vert\timestampseq\vert$ denotes the count of
time stamps and is mostly used as an upper bound for intervals of temporal
operators. 
The temporal structure $\tuple{\structseq, \timestampseq}$  satisfies formula $\varphi$
iff $(\structseq, \timestampseq, \emptyset, 0)\models_{CMFTBL} \varphi$.

\begin{table*}[t]
  \centering
  \begin{tabular}{lll} \smallskip
      $(\structseq, \timestampseq, v, i)\models_{CMFTBL} r(t_1,...,t_{a(r)}  )$ & iff & $( \beta(t_1,v, \struct_i),\cdots, \beta(t_{\arity(r)},v, \struct_i) ) \in r^{\struct_i}$ \\ \smallskip
      $(\structseq, \timestampseq, v, i)\models_{CMFTBL} (\lnot \psi)$ & iff & $(\structseq, \timestampseq, v, i)\not\models_{CMFTBL} \psi$\\  \smallskip
      $(\structseq, \timestampseq, v, i) \models_{CMFTBL} (\psi \lor \psi')$ & iff & $(\structseq, \timestampseq, v, i) \models_{CMFTBL} \psi$ or $(\structseq, \timestampseq, v, i) \models_{CMFTBL} \psi'$\\  \smallskip
      $(\structseq, \timestampseq, v, i) \models_{CMFTBL} (\exists x:\psi)$ & iff & $(\structseq, \timestampseq, v[x\mapsto d], i) \models_{CMFTBL} \psi$, for some $d\in \domain$\\
      $(\structseq, \timestampseq, v, i) \models_{CMFTBL} (\circ_I\psi)$ & iff & $\tau_{i+1}-\tau_i\in I$ and $(\structseq, \timestampseq, v, i+1)\models_{CMFTBL}\psi$\\  \smallskip
      & & and $(\structseq, \timestampseq, v, k)\models_{CMFTBL} \psi$, for all $k \in \mathbb{N}$ with $j<k\leq i$\\  \smallskip
      $(\structseq, \timestampseq, v, i)\models_{CMFTBL}(\psi U_I \psi')$ & iff & for some $j \geq i, \tau_j-\tau_i\in I, (\structseq, \timestampseq, v, j)\models_{CMFTBL} \psi'$,\\  \smallskip
      && and $(\structseq, \timestampseq, v, k)\models_{CMFTBL} \psi,$ for all $k\in\mathbb{N}$ with $i \leq k<j$\\  \smallskip
      $(\structseq, \timestampseq, v, i)\models_{CMFTBL} \nabla_I^p\psi$ & iff & 
      $(\structseq, \timestampseq, v, j) \models_{CMFTBL} \psi$,\\\smallskip
      && for at least fraction $p$ of indices $i\leq j\leq \vert\timestampseq\vert$ with 
      $\tau_j-\tau_i\in I$\\
      $(\structseq, \timestampseq, v, i)\models_{CMFTBL} (\downarrow^t_x\psi)$ & iff &  $(\structseq, \timestampseq, v[x\mapsto \beta(t,v,\struct_i)], i) \models_{CMFTBL} \psi$\\  \smallskip
  \end{tabular}
  \caption{Inductive definition of the relation $(\structseq, \timestampseq, v, i)\models_{CMFTBL} \psi$}
  \label{tab:cmftbl}
\end{table*}

For $I\in\mathbb{I}$ and the common Boolean constant $\top$ (for true), we define the usual syntactic shorthands and non-metric versions of operators as follows.
\begin{flushleft}
  \begin{tabular}{lcll}
      $(\varphi \land \psi)$ & $:=$ & $(\lnot((\lnot\varphi)\lor(\lnot\psi)))$ & logical and\\
      $(\varphi \Rightarrow \psi)$ & $:=$ & $((\lnot\varphi)\lor \psi)$ & implication\\
      $(\forall x:\varphi)$ & $:=$ & $(\lnot(\exists x:\lnot \varphi))$ & all quantifier\\
      $(\lozenge_I\varphi)$ & $:=$ & $(\top \; U_I \; \varphi)$  & eventually\\
      $(\square_I\varphi)$ & $:=$ & $(\lnot(\lozenge_I(\lnot\varphi)))$ & always\\
      $(\Delta^p_I\varphi)$ & $:=$ & $(\nabla_I^{1-p}\lnot \varphi)$ & max. prevalence\\
  \end{tabular}\\
\end{flushleft}
We obtain non-metric variants of the temporal operators for interval
$[0,\infty)$. The past-time operators from MFOTL (\emph{previously}, \emph{since}, \emph{once}, and
\emph{historically}) are not required in the study presented in this paper. They could
equally be defined for CMFTBL, but are omitted for brevity.

To enhance the readability (and also the writing) of CMFTBL formulas, we introduce several notational conventions.
Let $isVehicle \in \mathcal{R}$ be a unary relation.
We define the set of
all vehicles $\mathcal{V} \subseteq \domain{}$ as follows:
\begin{equation*}
\mathcal{V} := \{ d \in \domain \mid isVehicle(d) \}
\end{equation*}
Analogously, we define the set of pedestrians as $\mathcal{P}$, and the set of
actors $\mathcal{A} := \mathcal{P} \cup \mathcal{V}$ with $\mathcal{P} \cap \mathcal{V} = \emptyset$.
For our domain elements, we furthermore introduce a notation 
reminiscent of object-relational associations in programming
languages. For some vehicle $v \in
\mathcal{V}$ and the relations $\{isEgo, isLane, onLane\} \subseteq \mathcal{R}$ we use 
shorthand notations like:
\begin{equation*}
    \begin{split}
    v.isEgo &:= isEgo(v)\\
   v.lane &:= l \mid  l \in \domain \land isLane(l) \land onLane(v,l)
    \end{split}
\end{equation*}
All formulas used for the properties in our tree-based classifier need to be
evaluated for the ego vehicle and usually depend on one unary relation. Thus, for most formulas
$\varphi$ we can define a pattern for some $r \in \mathcal{R}$:
\begin{equation*}
    \varphi := \exists v \in \mathcal{V}: \square(v.isEgo) \land r(v) 
\end{equation*}
In such cases, we just define the relation $r$. For example, assuming $r =
obeyedSpeedLimit$, we could validate
that the ego vehicle at all times obeys the speed limit:
\begin{equation*}
    \begin{split}
        obeyedSpeed&Limit(v) :=\\
        &\ \square \big(v.speed \leq v.lane.speedLimitAt(v.pos)\big)
    \end{split}
\end{equation*}
The associations \emph{v.speed} and \emph{v.pos} are functions as
introduced before and \emph{speedLimitAt} is a
function from a position number and a lane to a speed limit number.
For numbers, we assume the relations $\set{eq, neq,
lt, gt, leq, geq} \in \rels$ to represent the common mathematical comparators
$\set{=, \neq, <, >, \leq, \geq}$, which we also allow as notation shortcuts.

With these notational conventions, we can quite straightforwardly define traffic
rules and environmental features using CMFTBL formulas and evaluate those
on sequences of scenes. Each predicate of our case study (cf.
Sect.~\ref{sec:evaluation}) is expressed this way. For comparison, consider the \emph{obeyedSpeedLimit}
formula without these syntactic conventions:
\begin{equation*}
    \begin{split}
        \varphi_{oSL} := \: & \exists v \in \domain: \square\big(isVehicle(v) \land isEgo(v)\big) \land \\
        & \square \Big(\exists l \in \domain: \exists p \in \domain: isLane(l) \land onLane(v,l) \land \\
        &   \land leq\big(speed(v), speedLimitAt(p, l)\big)\Big)
    \end{split}
\end{equation*}
Assumptions on the data can furthermore be validated
using dedicated formulas. For example, \emph{only one ego vehicle may exist at all times and it does
not change over time} could be expressed as:
\begin{equation*}
	\begin{split}
	   uniq&ueEgo := \\
	    & \exists v \in \mathcal{V}: \square \left(v.isEgo \land \forall v'\in\mathcal{V}: v'.isEgo \Rightarrow v = v' \right)
	\end{split}
\end{equation*}
Other data checks~--~e.g., that each vehicle can only be on one lane at a time or
that a vehicle's actual position on a lane can not be greater than the lane's length~--~can be added accordingly to ensure the single
relation nature of the object associations as well as overall data sanity and consistency.

\section{Classifiers for Scenarios and Metrics on Sets of Scenarios}
\label{sec:tscAndMetrics}

We want to use CMFTBL formulas for expressing 
features of scenarios and for classifying recorded
driving data into scenario classes.
Formally, we assume recorded driving data
to be given as temporal structures
$\tuple{\structseq, \timestampseq}$ over some 
fix basic signature. This basic signature is
the set of properties that is provided as 
information in the data, i.e., objects with 
positions and classifications on a road network with 
information about lanes, signs, and signals.

For the scope of this paper, we additionally assume
that the recorded data is already segmented into
sequences in a meaningful way. We use 
$\mathfrak{S}$ to denote a set of segments
of form $\tuple{\structseq, \timestampseq}$.
In practice, segmentation could either be done based
on a map or based on classification, e.g., of driving 
maneuvers of the ego vehicle, or by some other sensible
approach.\footnote{In our case study, we will segment data with the help of a map into 
sequences that contain one main driving situation:
driving through an intersection, driving along
a section of a multilane road between two intersections,
etc.}

We can then define classifiers that identify the scenario class
of some observed data $\tuple{\structseq, \timestampseq}$
and define metrics over observed classes of scenarios.

\subsection{Classifiers for Scenarios}
\label{subsec:scenario-classifiers}

Instead of simply using a set of features, we organize
features hierarchically in trees to account for dependencies
between features (a lane change, e.g., can only occur on 
a multi-lane road). This will enable us to capture the
taxonomies of features found in the 6-layer model
or in draft standards for specifying operational design 
domains.

\begin{definition}[Tree-Based Scenario Classifier]
  A tree-based scenario classifier (TSC) $\odd$ is a tuple
  $\tuple{\mathcal{Q}, q_r, \Gamma, \lambda_l, \lambda_u}$
  with
  \begin{itemize}
      \item set of nodes $\mathcal{Q}$ (i.e., the modeled features)
      \item root node $q_r \in \mathcal{Q}$
      \item set of edges $\Gamma$ of type $\grp{q, q', \varphi}$ where
      \begin{itemize}
          \item $q, q' \in \mathcal{Q}$ are source and destination, 
          \item CMFTBL formula $\varphi$ is the edge condition, 
      \end{itemize}  
      \item lower bounds for sub-features of nodes $\lambda_l: \mathcal{Q} \to \mathbb{N}_0$
      \item upper bounds for sub-features of nodes $\lambda_u: \mathcal{Q} \to \mathbb{N}_0$
  \end{itemize}  
\end{definition}
We write $q \xrightarrow{\varphi} q'$ for $\grp{q, q', \varphi}$.
We require $\odd$ to be a tree rooted at $q_r$.
For $q\in \mathcal{Q}$, let $c(q) = \set{q' ~|~ q \xrightarrow{\varphi} q' 
  \in \Gamma}$ denote the children of $q$.
Bounds must be $ 0 \leq \lambda_l(q) \leq \lambda_u(q) \leq |c(q)|$.
A path of length $k$ in $\odd$ is a sequence of $k$ transitions $q_{i-1}
\xrightarrow{\varphi_i} q_{i}$ with $1 \leq i \leq k$ and $q_0 = q_r$.

Inspired by feature models, we name certain types of nodes $q \in \mathcal{Q}$ 
depending on their lower and upper bounds (abbreviated notation with parentheses):
\begin{description}[labelindent=2\parindent,labelwidth=9em]
  \item[\textbf{All / (A)}] $\lambda_l(q) = \lambda_u(q) = |c(q)|$
  \item[\textbf{Exclusive / (X)}] $\lambda_l(q) = \lambda_u(q) = 1$
  \item[\textbf{Optional / (O)}] $\lambda_l(q) = 0 \land \lambda_u(q) = |c(q)|$
  \item[\textbf{a/b-Bounded / (a..b)}] $\lambda_l(q) = a \land \lambda_u(q) = b$
  \item[\textbf{Leaf / ( )}] $\lambda_l(q) = \lambda_u(q) = 0$
\end{description}
We introduce bounds on sub-features as a means of computing an upper bound
on the number of combinatorial combinations, i.e., the number of observable 
scenario classes, in the next section. 
A more precise approach to computing feasible scenarios would be to 
compute satisfiable combinations of features. Such an approach, 
however, does not seem feasible or meaningful.
Even if the satisfiability of some fragment of CMFTBL can be established,
there is no mechanism to constrain acceptable models to realistic segments.

We can now describe individual scenario classes for a scenario classifier.

\begin{definition}[Scenario Class]
For a given tree-based scenario classifier  
$\odd = \tuple{\mathcal{Q}^o q_r^o, \Gamma^o, \lambda_l^o, \lambda_u^o}$, 
a scenario class is a tree $T= \tuple{\mathcal{Q}, q_r, \Gamma}$ 
with 
\begin{itemize}
   \item set of nodes $\mathcal{Q} \subseteq \mathcal{Q}^o$
   \item root node $q_r = q_r^o$
   \item  set of edges $\Gamma$ of type $\grp{q, q'}$ and such that $\grp{q, q', \varphi} \in \Gamma^o$
\end{itemize}
We require the number of children $c(q)$
for every node $q \in \mathcal{Q}$ to be within the lower and upper bounds of $q$ in $\odd$.
\label{def:scenario-class}
\end{definition}

Let $\tscclasses_\odd$ denote the (finite) set of all scenario 
classes for tree-based classifier $\odd$,
and let $\mathcal{W}$ denote the (infinite) set of observable
segments of driving data $\tuple{\structseq, \timestampseq}$.
We denote the classification function that maps observed driving data 
$\tuple{\structseq, \timestampseq}$ to a scenario class $T$ based on 
the tree-based scenario classifier $\odd$ 
by $C_{\odd}:\mathcal{W} \to \tscclasses_\odd$.
For recorded data segment $\mathcal{S}=\tuple{\structseq, \timestampseq}$, 
we compute $C_{\odd}(\mathcal{S}) = \tuple{\mathcal{Q},
q_r, \Gamma}$ by computing the set $\mathcal{Q}$ of nodes, 
which uniquely determines the set of transitions. 
We initialize $\mathcal{Q}$ as $\set{q_r^o}$ and then
add every node $q'$ for which 
$q\in\mathcal{Q}$ and $\grp{q, q', \varphi} \in \Gamma^o$ 
with $\mathcal{S}\models \varphi$ until a fixed point is reached.
We assume that bounds permit that a valid class 
can be computed for every realistic segment $\mathcal{S}$
and lift $C_{\odd}$ to sets of segments by letting 
$C_{\odd}(\scenarios)$ denote the set of observed scenario 
classes for \scenarios{}.

\subsection{Coverage Metrics for Sets of Scenarios} 
\label{sec:coverage}

Given a set \scenarios{} of recorded segments and a classifier $\odd$, 
we want to analyze and quantify \emph{if and to which degree} the
recorded data covers possible scenarios. 

We start by showing how to compute the number of scenario classes for
a tree-based scenario classifier $\odd = \tuple{\mathcal{Q}, q_r, \Gamma, \lambda_l, \lambda_u}$.
Let $\Gamma_q = \{ (q,q',\varphi)~\in~\Gamma \}$ be the set of edges
originating in $q$, and
$$[\Gamma_q]^{\lambda_l(q)..\lambda_u(q)} ~ =_{def} ~ \bigcup_{i = \lambda_l(q)}^{\lambda_u(q)} [\Gamma_q]^i$$ be the set of all
subsets of these edges with size within lower bound and upper bound of $q$.
We define the size $|\odd|$ = $|q_r|$
recursively as
\begin{equation*}
|q|  ~ =_{def} ~ \sum_{G \in [\Gamma_q]^{\lambda_l(q)..\lambda_u(q)} } \left ( \prod_{(q,q',\varphi) \in G} | q' |  \right )\mbox{.}
\end{equation*}
The primary metric we are considering is \textbf{scenario class coverage} (SCC),
expressing the ratio between the amount of observed scenario classes and 
the number of classes modeled by classifier
\odd{} = $\tuple{\mathcal{Q}, q_r, \Gamma, \lambda_l, \lambda_u}$.  
For a set \scenarios{} of recorded segments, we define
\begin{equation*}
\textrm{SCC}(\scenarios, \odd) ~ =_{def} ~  \frac{|  \mathcal{C}_{\odd}(\scenarios) |}{|\odd|}
\end{equation*}

It can be expected that gaining high coverage on TSCs with (potentially
multiple combinations of) rare events requires an increasingly high amount of
test scenarios. To measure the individual rarity of the modeled environmental
conditions, from which explanations for coverage gaps might be derived, we
introduce a metric for  \textbf{absolute feature occurrence} (afo). It
counts the number of segments 
that are classified as scenarios containing a given node (i.e., feature). 
\begin{equation*}
  \textrm{afo}(\scenarios, q) ~ =_{def} ~ | \{  \tuple{\mathcal{Q}, q_r, \Gamma} \in 
  \mathcal{C}_{\odd}(\scenarios) ~ | ~ q \in \mathcal{Q}\} |
\end{equation*}
In addition to coverage, which only considers if a scenario class
has been observed, we define \textbf{scenario instance count} (sic) to count 
how often a certain class has been encountered in a set of scenarios.
\begin{equation*}
  \textrm{sic}(\scenarios, t) ~ =_{def} ~ | \set{ \mathcal{S} \in \scenarios ~ | ~ \mathcal{C}_{\odd}(\mathcal{S}) = t } |
\end{equation*}
Similar to calculating the size of a TSC, we can enumerate all
possible scenario classes and use them to identify \textbf{class instance
missings}, i.e., classes as which no $\scenario \in \scenarios$ is classified.
However, gaining meaningful insights from large sets of missing classes is
difficult. Therefore, we also analyze \textbf{feature pair misses}, i.e., pairs
of TSC nodes that do not exist together in any observed class.

\forestset{
    ew edges new/.style={
        for tree={
            l sep+=50pt
        }
    },
}
\begin{figure*}
    \resizebox{0.99\textwidth}{!}{
        \begin{forest}
            for tree={
                grow=east,
                s sep-=8pt,
                l sep+=35pt
            }
            [Root (A) \fullallcircle \halfstaticcircle \halfdynamiccircle \halfdrivingcircle \halfenvcircle \halfpedestriancircle
            [Time of Day (X) \fullenvcircle \fullpedestriancircle, edge label={node[midway,fill=white,font=\scriptsize]{$\top$}}
            [Noon]
            [Sunset]
            ]
            [Traffic Density (X) \fulldynamiccircle \fulldrivingcircle \fullenvcircle, edge label={node[midway,fill=white,font=\scriptsize]{$\top$}}
            [Low]
            [Middle]
            [High]
            ]
            [Road Type (X) \halfstaticcircle \halfdynamiccircle \halfdrivingcircle \halfpedestriancircle, edge label={node[midway,fill=white,font=\scriptsize]{$\top$}}    
            [Multi-Lane (A) \halfstaticcircle \halfdynamiccircle \halfdrivingcircle \halfpedestriancircle, edge label={node[midway,fill=white,font=\scriptsize]{(\ref{eq:multiLane})}}
            [Stop Type $(0..1)$ \fullstaticcircle \fulldrivingcircle, edge label={node[midway,fill=white,font=\scriptsize]{$\top$}}
            [Has Rel. Red Light]
            ]
            [Maneuver (X) \fullstaticcircle \fulldrivingcircle, edge label={node[midway,fill=white,font=\scriptsize]{$\top$}}
            [Lane Follow, edge label={node[xshift=-50pt, yshift=2pt, fill=white, font=\scriptsize]{$\lnot$(\ref{eq:changedLane})}}]
            [Lane Change, edge label={node[xshift=-50pt, yshift=-2pt, fill=white, font=\scriptsize]{(\ref{eq:changedLane})}}]
            ]
            [Dynamic Relation (O) \fulldynamiccircle \fulldrivingcircle \halfpedestriancircle, edge label={node[midway,fill=white,font=\scriptsize]{$\top$}}
            [Pedestrian Crossed \fullpedestriancircle, edge label={node[midway,fill=white,font=\scriptsize]{(\ref{eq:pedestrianCrossed})}}]
            [Overtaking]
            [Oncoming Traffic]
            [Following Leading Vehicle]
            ]
            ]        
            [Single-Lane (A) \halfstaticcircle \halfdynamiccircle \halfdrivingcircle \halfpedestriancircle, edge label={node[midway,fill=white,font=\scriptsize]{(\ref{eq:singleLane})}}
            [Dynamic Relation (O) \fulldynamiccircle \fulldrivingcircle \halfpedestriancircle, edge label={node[midway,fill=white,font=\scriptsize]{$\top$}}
            [Pedestrian Crossed \fullpedestriancircle, edge label={node[midway,fill=white,font=\scriptsize]{(\ref{eq:pedestrianCrossed})}}]
            [Oncoming Traffic]
            [Following Leading Vehicle]
            ]
            [Stop Type $(0..1)$ \fullstaticcircle \fulldrivingcircle
            [Has Rel. Red Light]
            [Has Yield Sign]
            [Has Stop Sign]
            ]    
            ]
            [Junction (A) \halfstaticcircle \halfdynamiccircle \halfdrivingcircle \halfpedestriancircle, edge label={node[midway,fill=white,font=\scriptsize]{(\ref{eq:junction})}}
            [Maneuver (X) \fullstaticcircle \fulldrivingcircle
            [Left Turn]
            [Right Turn]
            [Lane Follow]
            ]
            [Dynamic Relation (O) \fulldynamiccircle \fulldrivingcircle \halfpedestriancircle
            [Must Yield]
            [Pedestrian Crossed \fullpedestriancircle]
            [Following Leading Vehicle]
            ]
            ]
            ]
            [Weather (X) \fullenvcircle \fullpedestriancircle, edge label={node[midway,fill=white,font=\scriptsize]{$\top$}} 
            [Clear]
            [Cloudy]
            [Wet]
            [Wet Cloudy]
            [Soft Rain]
            [Mid Rain]
            [Hard Rain]
            ]
            ]
        \end{forest}
    }
    \caption{
        This figure shows the full classifier structure used for the analyses in this paper. Hereby, edge labels reference the 
        related logical formula that is deciding whether an edge is taken. We use the following classifier sets:\\
        \halfallcircle\ \emph{full TSC}, 
        \halfstaticcircle\ \emph{layer 1+2}, 
        \halfdynamiccircle\ \emph{layer 4}, 
        \halfdrivingcircle\ \emph{layer 1+2+4}, 
        \halfenvcircle\ \emph{layer (4)+5} and 
        \halfpedestriancircle\ \emph{pedestrian}.
    }
    \label{fig:tsc}
\end{figure*}
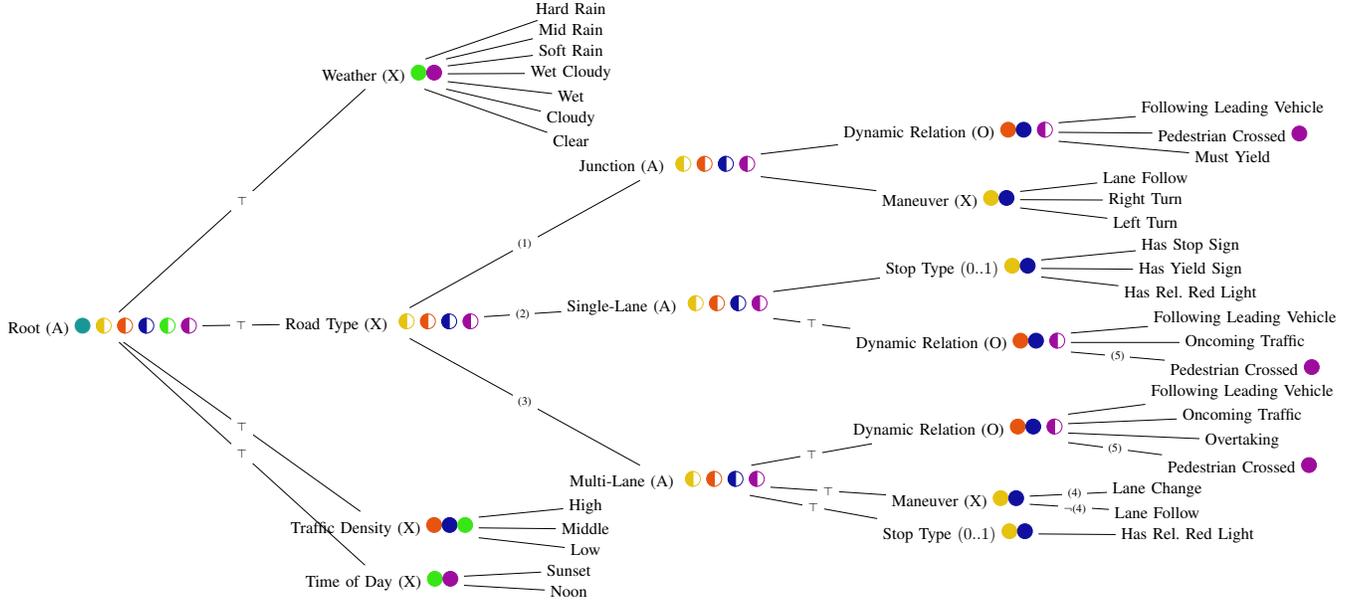
  
\section{Evaluation}
\label{sec:evaluation}

Our evaluation is designed as a single case mechanism experiment~\cite{Wieringa2014SingleCaseMechanism}
that validates the presented approach and our implementation.
We develop a tree-based scenario classifier for an urban 
driving environment and use it for analyzing simulated urban 
traffic. Features are chosen to model 
the types of properties (or labels) that are envisioned for 
specifications of operational design domains (ODDs) as
described in BSI 1883~\cite{BSI2020OperationalDesignDomain} or OpenODD~\cite{StandardizationofAutomation2021AsamOpenoddConcept}.

Since this is the first work on scenario class coverage, we aim at answering the following questions ---
mostly with qualitative data.
\begin{enumerate}
\item[Q1.] Can relevant properties of operational design domains
           be expressed in CMFTBL? 
\item[Q2.] Is it computationally feasible to classify scenarios with 
           a tree-based scenario classifier?
\item[Q3.] To which degree can scenario class coverage be achieved and
           can scenario class coverage generate useful insights (e.g.,
           missing classes)?
\end{enumerate}
The remainder of this section discusses the 
classifier developed for our case study, details the experimental setup, presents
results from the simulated experiments, and provides
initial answers to the above questions.

\subsection{Tree-Based Scenario Classifier Definition}
\label{subsec:sct-def}
To construct a tree-based scenario classifier (TSC) for our case study, we 
evaluated the 6-layer model of scenario classification by Scholtes 
et al.~\cite{Scholtes20216LayerModel} and extracted observable 
features. We defined logical formulas using CMFTBL
that are capable of identifying these features on segments (i.e., sequences of scenes). 
The hierarchic organization of all the features resulted in the TSC visualized in Fig.~\ref{fig:tsc}.
We additionally define smaller TSCs by
grouping features that we want to analyze together.  
This allows us to study coverage 
for smaller sets of features.
For the sake of
presentation in this paper, we introduce TSC projections. They combine related
features into subsets of all features of
an original \emph{full TSC}. The following projections are based on the layers of
information discussed in~\cite{Scholtes20216LayerModel}:
\begin{itemize}
  \item \textbf{full TSC}: the complete TSC that serves as the reference point for the comparison with the other projections
  \item \textbf{layer 1+2}: driving features in relation to static information (roads, lanes traffic signs, etc.) during the scenario run
  \item \textbf{layer 4}: driving features in relation to other objects that dynamically change during the scenario run (like other vehicles, pedestrians, etc.)
  \item \textbf{layer 1+2+4}: combination of static information and dynamically changing objects
  \item \textbf{layer (4)+5}: environmental features from layer 5 combined with the traffic density from layer 4
  \item \textbf{pedestrian}: example for a more specialized projection to analyze the coverage of pedestrians crossing the street in all possible environmental situations
\end{itemize}
We did not include Layer 3 \emph{(Temporary Modifications of Layer 1 and Layer 2)}
and Layer 6 \emph{(Digital Information)}, as there were no elements
of these layers available in our test environment. 

We visualize our projection labels in
Fig.~\ref{fig:tsc} through colored circles. A filled circle \fullcirc\
indicates that the complete subtree rooted at this node is included in the
projection. 
Half circles \halfcirc\ indicate that the subtree rooted at this node is 
partially (i.e., as labeled) included in the projection. 
All edges of the TSC have a corresponding logical condition.
For readability reasons, Fig.~\ref{fig:tsc} only depicts \emph{always-true} edge conditions and
formulas explicitly described in this paper.
A more detailed overview of the set of implemented predicates is given in the next section.
There, we discuss the total number of predicates, their definitions using our newly introduced operators, and their
complexity.

\subsection{Predicates}
\label{subsec:predicates}
For our case study, we defined all scenario class features as CMFTBL predicates and formulas. This section discusses a selection of these
predicates, in particular to demonstrate our newly introduced prevalence and binding CMFTBL operators.

Let $\mathcal{V}$ be the set of all vehicles and $\mathcal{P}$ be the set of 
all pedestrians. As a basis for our data structure, we use the OpenDrive standard\footnote{\url{https://www.asam.net/standards/detail/opendrive/}}.
Therefore, we can reason about static and dynamic relations between vehicles
and other entities (e.g., other vehicles, pedestrians, landmarks, etc.) by
mapping entity positions to their respective OpenDrive lanes. Consequently,
each entity is related to a lane which in itself contains additional information
we use for our predicate definitions.

To decide whether a vehicle $v \in \mathcal{V}$ was primarily driving through a junction during the analyzed segment,
we require the vehicles' road to be categorized as a junction in at least 80\% 
of the time stamps.
\begin{equation}
  \label{eq:junction}
  \begin{split}
      isInJunction(v) := &\ \nabla^{0.8}(v.lane.road.isJunction)
  \end{split}
\end{equation}
Similarly, to determine whether a vehicle $v \in \mathcal{V}$ is driving on a single-lane road, 
we require $v$'s road to have only one lane for at least 80\% of the observed scenes.
Additionally, we require the road to not be classified as a junction.
\begin{equation}
  \label{eq:singleLane}
  \begin{split}
      onSingleLa&neRoad(v) := \lnot isInJunction(v) \land\\
      &\nabla^{0.8}(sameDirectionLaneCount(v.lane) = 1)
  \end{split}\raisetag{2\baselineskip}
\end{equation}
We define the predicate for deciding if a vehicle $v \in \mathcal{V}$ is on a multi-lane road by
combining the predicates (\ref{eq:junction}) and (\ref{eq:singleLane}).
\begin{equation}
  \label{eq:multiLane}
  \begin{split}
      onMultiLaneRoad(v) :=         & \lnot onSingleLaneRoad(v)\land\\
      & \lnot isInJunction(v)
  \end{split}
\end{equation}   
To be able to detect a lane change for a given vehicle 
$v \in \mathcal{V}$, our binding operator is utilized.
We bind the lane of vehicle $v$ at the first evaluation
time stamp to a new variable $l$. As the vehicle $v$ progresses in time
and might change its lane, we can compare its lane value to $l$
to detect a lane change.
\begin{equation}
  \label{eq:changedLane}
  \begin{split}
      changedLane(v) := & \bind{v.lane}{l}\ \big(\lozenge(l \neq v.lane)\big)
  \end{split}
\end{equation}   
Pedestrians crossing a lane are (at some timestamp) identified as being \emph{on} this lane.
Therefore, we can detect if for a pedestrian $p \in \mathcal{P}$ 
and vehicle $v \in \mathcal{V}$ the predicate \code{onSameLane(v,p)} holds. 
For the vehicle $v$, a crossing of pedestrian $p$ is only relevant if it happens `closely in front of $v$'. 
This is defined by \code{inReach(p,v)} which checks if $p$'s position on the (same) lane is in front of and at most $10$ meters away from $v$.
\begin{equation}
  \label{eq:pedestrianCrossed}
  \begin{split}
      pedest&rianCrossed(v) := \\
      & \lozenge \big(\exists p \in \mathcal{P}:  onSameLane(p, v) \land inReach(p, v)\big)\\
      onSa&meLane(a_0, a_1) := a_0.lane = a_1.lane\\
      &inReach(a_0, a_1) :=  0 \leq a_0.pos - a_1.pos \leq 10 
  \end{split}\raisetag{2\baselineskip}
\end{equation}
All predicates defined for the tree-based scenario classifier in Fig.~\ref{fig:tsc} are of similar complexity as the ones presented above. 
In total, we defined $51$ predicates (including sub-predicates) to completely express the detection of the modeled features for our experiments.
We used the min. prevalence operator in $18$ predicates to model that some 
feature is present for most of the time covered by a segment.
The binding operator was used once directly, for specifying the change of lanes,
and once indirectly by using the negation of the change of lanes.

\begin{figure*}
    \begin{subfigure}{0.48\linewidth}
        \centering                
        \includegraphics[height=4.5cm]{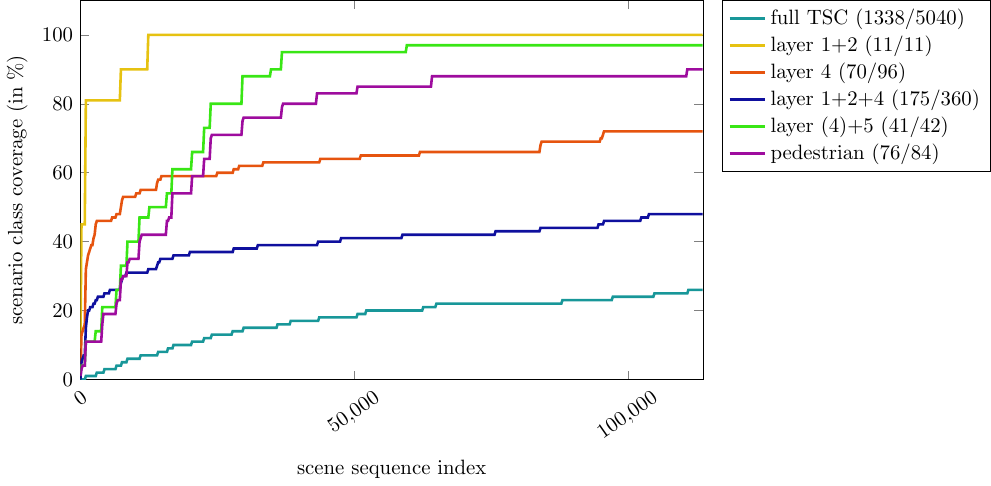}
        \caption{Scenario class coverage over the course of the 113,767 analyzed segments}
        \label{subfig:coveragePercent}
    \end{subfigure}\hfill
    \begin{subfigure}{0.48\linewidth}
        \centering
        \includegraphics[height=4.5cm]{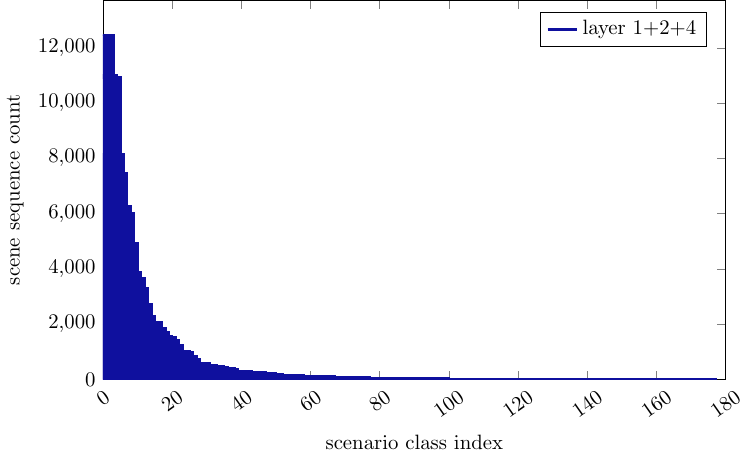}
        \caption{Distribution of the 175 observed scenario classes for the \emph{layer 1+2+4} projection}
        \label{subfig:drivingDistribution}
    \end{subfigure}
    \caption{Coverage of scenario classes and distribution of segments over classes}
    \label{fig:plots}
\end{figure*}

\subsection{Experimental Setup}

As the basis for our experiments, we built a toolchain using the CARLA
simulator~\cite{Dosovitskiy2017CarlaOpenUrban} and an analysis framework written
in the Kotlin programming language\footnote{\url{https://kotlinlang.org}}
that classifies recorded scenario runs according to a tree-based classifier.
As a proof-of-concept implementation, it is not optimized for performance.
However, it is already sufficient
to run our experiments within a few hours. Thus, we left a proper algorithmic
approach to CMFTBL evaluation for future work.

Based on the classifications of recorded runs, a subsequent analysis step calculates the
coverages and analyses introduced in Sect.~\ref{sec:coverage}. Additionally, by
iteratively analyzing the set of recorded segments, we can measure class
coverage over time by counting newly observed scenario classes.
This provides us with an increasing curve on coverage.

In our toolchain, TSCs are evaluated on an abstract representation of the scene,
i.e., the ego vehicle and its surroundings. This data structure
is designed to be constructed in various ways, and we provide an
implementation for CARLA. Static as well as dynamic data is
exported into JSON files during simulation, read by the Kotlin
framework, and weaved together forming the consistent abstract view of the world.

The data for each simulation run is then segmented to be classified. The
primary factor for this segmentation is the ego vehicles' road.
This results in each simulation run being cut into individual segments of
either driving through a junction or following a (potentially multi-lane) road section
without crossed lanes. After this segmentation, we discard all segments too short for a meaningful analysis, i.e., segments containing only 10 or fewer scenes.

For our experiments, we recorded $100$ simulation runs of $5$ minutes each.  In
every run, a random map, daytime, and weather were chosen and up to $200$
vehicles and $30$ pedestrians were spawned randomly on the map. For maps that do
not specify enough spawn points, we spawned as many actors as possible.

During the simulation, all vehicles drove around the map using CARLA's autopilot. 
We analyzed each simulation run multiple times: once with each vehicle
being considered to be \emph{the ego vehicle}. This enabled us to increase the
amount of encountered situations (and therefore coverage) without the need to
record about $200$ times as many simulation runs.  Overall, this resulted in
$113,767$ analyzed segments representing $1,104$ hours of driving data. The
analysis of this data with the classifiers and predicates introduced in
Sects.~\ref{subsec:sct-def} and~\ref{subsec:predicates} takes about $118$
minutes on a single core of a 2021 Apple M1 Pro SoC.  A reproduction package
for our experiments~--~a virtual machine image that contains our recorded
driving data, the framework, specifications, and analysis code~--~is
archived on Zenodo~\cite{Schallau2023}.

\subsection{Experimental Results}

In this section, we present the application of our coverage metrics and analyses for scenario classes based on our data set of $113,767$ classified segments.

\medskip\noindent\textbf{Class Coverage.}
We visualize our results for scenario class coverage over the course of analyzed segments in Fig.~\ref{subfig:coveragePercent}. Each
colored curve represents the coverage result of one classifier projection, as defined in Sect.~\ref{subsec:sct-def}.
The legend also shows for each projection the final count of observed classes after analysis of all $113,767$ segments as well as the number of possible classes.
The \emph{layer 1+2} projection covers $100\%$ of scenario classes after $12,233$ analyzed segments.
Furthermore, \emph{layer (4)+5} almost fully covers the possible scenario classes after around $59,409$ segments,
but misses one scenario class and therefore only reaches $97\%$. The \emph{pedestrian} projection
covers over $90\%$ of relevant scenario classes. The projections \emph{layer 4} and 
\emph{layer 1+2+4} cover $72\%$ and $48\%$ of relevant scenario classes, respectively.
Finally, the reference projection \emph{full TSC} reaches a coverage of $26\%$.

\medskip\noindent\textbf{Scenario Instance Count.}
In Fig.~\ref{subfig:drivingDistribution}, we exemplarily demonstrate the scenario instance count metric
of the $175$ observed scenario classes for the \emph{layer 1+2+4} projection. The plot
shows a long-tail distribution in which $85,120$ segments 
of the total $113,767$ segments
are classified into only $15$ scenario classes. 
The remaining $28,647$ segments are classified into the remaining
$160$ scenario classes. The three most common scenario classes are each observed about $11,000$ times.

\begin{table*}
\centering
\begin{tabular}{l|c|c|c|c|c|c|c|c|c|c|c|c|c}
	\multicolumn{1}{c|}{Map} & Data [h] & Segments [\#] & Road Sections [\#] & \multicolumn{2}{|c|}{Seg./R.S. [\#]} & \multicolumn{4}{|c|}{Seg. length [\#] } & \multicolumn{4}{|c}{Lane length [m]} \\ 
	& & & & AVG & SD & Avg & SD & Min & Max & Avg & SD & Min & Max\\
		\hline
    Town 01 & $467.4$ & $38,190$ & $38$ & $1,005.0$ & $696.8$ & $89.1$ & $68.1$ &  $11$ & $580$ & $52.7$ & $68.8$ & $16$ & $310$ \\
    Town 02 & $357.1$ & $38,702$ & $28$ & $1,382.2$ & $856.4$ & $67.4$ & $51.8$ & $11$ & $592$ & $34.2$ & $32.5$ & $16$ & $178$\\
    Town 10 & $279.7$ & $36,875$ & $32$ & $1,152.3$ & $815.7$ & $55.6$ & $69.7$ & $11$ & $584$ & $36.5$ & $27.0$ & $4$ & $122$ \\
    \hline
    Total & $1,104.2$ & $113,767$ & $98$ \\
\end{tabular}
\caption{Analysis of simulated test drives per map: driving time, road sections, segments per road sections, segment lengths, and lane lengths}
\label{tab:meta}
\end{table*}

\medskip\noindent\textbf{Test Scenario Set Analysis.}
Table~\ref{tab:meta} gives an overview of the statistical values of our generated test scenario set. 
We used three maps shipped with the CARLA simulator with an average lane length of $37.4$~meters. Note that
especially \emph{Town 01} has some long lane sections with the maximum length being $310$~meters. In contrast,
some lane sections are only $4$~meters long. In total, we generated $1,104.2$ hours of data with a 
segment count of $113,767$. On average, there are over $1,000$ segments per road section of each map, while
the segments have an average length of over $50$ scenes with a maximum scene count of $592$.

\begin{figure*}
    \begin{subfigure}{0.39\textwidth}
        \begin{subfigure}{0.49\textwidth}
            \includegraphics[height=2.75cm]{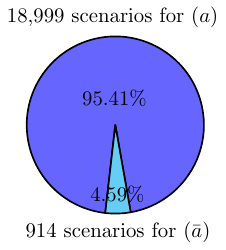}
        \end{subfigure}\hfill
        \begin{subfigure}{0.49\textwidth}
            \includegraphics[height=2.75cm]{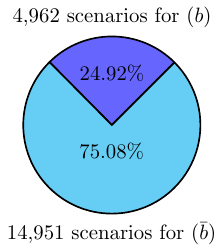}
        \end{subfigure}
        \begin{subfigure}{0.49\textwidth}
            \includegraphics[height=2.75cm]{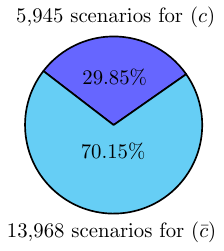}
        \end{subfigure}\hfill
        \begin{subfigure}{0.49\textwidth}
            \includegraphics[height=2.75cm]{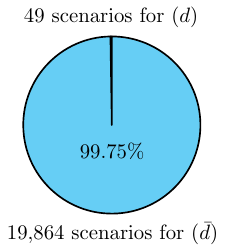}
        \end{subfigure}
        \caption{Distributions of individual feature occurrences}
        \label{subfig:single-pie-charts}
    \end{subfigure}
    \begin{subfigure}{0.6\textwidth}
        \centering                
        \includegraphics[width=\textwidth]{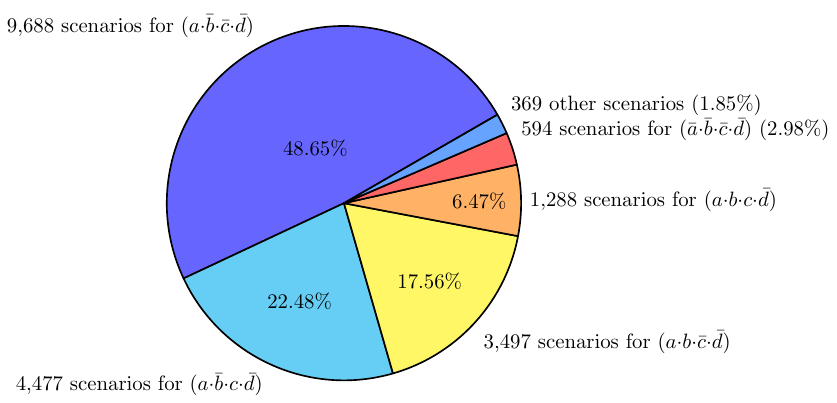}
        \caption{Distribution of all feature combinations}
        \label{subfig:combined-pie-chart}
    \end{subfigure}
    \caption{Distributions of ``dynamic relations'' for ``multi-lane'' roads. We define \emph{a}=``Oncoming Traffic'', \emph{b}=``Pedestrian Crossed'', \emph{c}=``Following Leading Vehicle'' and \emph{d}=``Overtaking''. The 369 other scenarios are composed of the following combinations: 152 scenarios for ($\bar{a}{\cdot}\bar{b}{\cdot}c{\cdot}\bar{d}$), 143 scenarios for
        ($\bar{a}{\cdot}b{\cdot}\bar{c}{\cdot}\bar{d}$),  37 scenarios for
        ($a{\cdot}\bar{b}{\cdot}\bar{c}{\cdot}d$), 25 scenarios for
        ($\bar{a}{\cdot}b{\cdot}c{\cdot}\bar{d}$), 9 scenarios for
        ($a{\cdot}b{\cdot}\bar{c}{\cdot}d$) and 3 scenarios for 
        ($a{\cdot}\bar{b}{\cdot}c{\cdot}d$). The remaining combinations
        ($a{\cdot}b{\cdot}c{\cdot}d$),
        ($\bar{a}{\cdot}b{\cdot}c{\cdot}d$)
        ($\bar{a}{\cdot}b{\cdot}\bar{c}{\cdot}d$),
        ($\bar{a}{\cdot}\bar{b}{\cdot}c{\cdot}d$) and
        ($\bar{a}{\cdot}\bar{b}{\cdot}\bar{c}{\cdot}d$) never occurred.}
    \label{fig:pieChartCombined}
\end{figure*}

\medskip\noindent\textbf{Absolute Feature Occurrence.}
Our analysis provides detailed insights into specific scenario classes regarding the underlying features and their combinations. To demonstrate
the results, Fig.\ref{fig:pieChartCombined} exemplarily shows analyses on the \emph{Dynamic Relation} features of the \emph{Multi-Lane} node of our
TSC (cf. Fig.~\ref{fig:tsc}). For better readability in the figure, we label the observable features as 
\emph{a}=``\emph{Oncoming Traffic}'', \emph{b}=``\emph{Pedestrian Crossed}'', \emph{c}=``\emph{Following Leading Vehicle}'' and \emph{d}=``\emph{Overtaking}''.
We also write $(x)$ or $(\bar{x})$ if feature \emph{x} was observed or not observed, respectively.
For example, the combination $(a{\cdot}b{\cdot}c{\cdot}\bar{d})$ describes the scenario classes in which 
\emph{Oncoming Traffic}, \emph{Pedestrian Crossed} and \emph{Following Leading Vehicle} are observed, while 
\emph{Overtaking} is not observed. 
Figure~\ref{subfig:single-pie-charts} visualizes the individual absolute occurrences of each observable feature for the dynamic relations on multi-lane roads.
The percentages are based on the $19,913$ analyzed segments classified as containing the \emph{Multi-Lane} feature.
Here, \emph{Oncoming Traffic}~(a) appears in $95.41\%$ of the total occurrences. \emph{Pedestrian Crossed}~(b) and \emph{Following Leading Vehicle}~(c) are similarly
present with a coverage of $24.92\%$ and $29.85\%$, whereas \emph{Overtaking}~(d) is only encountered in $0.25\%$ of the analyzed segments.

\medskip\noindent\textbf{Full Combinatorial Analysis.}
As defined in our TSC (cf. Sect.~\ref{subsec:sct-def}) the \emph{Dynamic Relation} node is 
marked as \emph{Optional}, i.e., all combinations of the four children form valid scenario classes. Consequently, there are $16$ possible combinations of features.
Figure~\ref{subfig:combined-pie-chart} visualizes the distribution for all combinations of
features. It can be seen that $95.16\%$ of observed scenarios are covered by the following
four feature combinations: $(a{\cdot}\bar{b}{\cdot}\bar{c}{\cdot}\bar{d})$, $(a{\cdot}\bar{b}{\cdot}c{\cdot}\bar{d})$, $(a{\cdot}b{\cdot}\bar{c}{\cdot}\bar{d})$ and $(a{\cdot}b{\cdot}c{\cdot}\bar{d})$.
Of the five feature combinations that never occurred, each includes feature $(d)$, which directly stems from the overall low occurrence of only $0.25\%$ of feature $(d)$.
Additionally, the three observed combinations that include feature $(d)$ are among the four combinations with lowest occurrence.

\medskip\noindent\textbf{Feature Pair Misses.}
As discussed before, our method yields precise information on which scenario
classes never occurred. But as the full TSC analysis resulted in 3,702 unseen
classes, a detailed analysis is unfeasible.  With the analysis of feature pair
misses, we instead focus on predicate combinations that never occurred.  This
results in the information that the following five predicate combinations were
never observed together: (Overtaking \& Lane Change), (Overtaking \& Has Red
Light), (Has Stop Sign \& High Traffic), (Has Yield Sign \& High Traffic), (Has
Yield Sign \& Middle Traffic).

\subsection{Discussion}
\label{subsec:discussion}

In the previous section, we visualized and described various methods of analyzing test drives
regarding a given specification. We demonstrated the expressiveness of our approach
with coverage metrics for scenario classes and predicate combinations. This
section discusses these findings in the context of the three questions
formulated on page \pageref{sec:evaluation} and closes with a discussion on
threats to validity.

\medskip\noindent\textbf{Q1 (Expressivity).}
Using the CMFTBL logic, we were able to express many relevant properties for
common driving situations considered in the proposals for operational design
domains~\cite{BSI2020OperationalDesignDomain} and approaches like the 6-layer model~\cite{Scholtes20216LayerModel}. 
In particular, the prevalence operator was helpful to detect properties where
it is natural to formulate `majority of the time' constraints (like
environmental conditions or traffic density). The binding operator
adds an intuitive mechanism for value storage that can be used to include
`remembered information from the past' in the evaluation of a state.
The notational conventions (like dedicated sets for vehicles/pedestrians or
object-relational element associations)  furthermore facilitate a mapping from
a more human-readable presentation to CMFTBL formula syntax.
Properties we did not include in our case study were usually left out not
because it was impossible (or even particularly inconvenient) to be expressed using
CMFTBL, but because we were not able to automatically extract~--~with a reasonable
amount of effort~--~the required information from our simulation setup with
CARLA (e.g., yield priorities in roundabouts, behavior on highway entries, or
temporary modifications like construction work).
We are  confident that the logic can express most of the features
required of a scenario classifier for an ODD.

\medskip\noindent\textbf{Q2 (Analysis Time).}
We analyzed a total of $1,104$ hours of data from simulated test drives, which
took a little over $118$ minutes. With a total of $113,767$ segments we have on
average $34.93$ seconds of driving data per segment and $62.23$ milliseconds of
computation time per segment evaluation. 
While a more comprehensive scenario classifier would contain 
more features, due to the tree-based structure of our classifier, 
whole sub-trees get cut off from evaluation if a condition does not hold 
(e.g., none of the \emph{single-lane} features of Fig.~\ref{fig:tsc} are 
evaluated when the segment is recognized as a \emph{junction}).
The obtained results thus indicate that our approach is
generally feasible with regard to computation time. Even online monitoring of 
properties while driving (i.e., after completing a segment) seems possible. 

\medskip\noindent\textbf{Q3 (Scenario Coverage).}
Our experiments demonstrate that scenario coverage can be achieved with
our concept of hierarchical classifiers. Even though the features evaluated with
our classifiers are limited in scope, they cover a sensible amount of 
situations for urban driving. 
With our approach, it is possible to 
automatically classify test drives
based on a predefined specification. 
Our detailed analyses proved particularly useful for the interpretation of the
coverage levels our projections converge to. All five feature combinations not
encountered at all throughout our data combine a \emph{layer 1+2} feature with
a \emph{layer 4} feature. Due to the combinatorial nature of our classifier
concept, about half of all combinations in the \emph{layer 1+2+4} projection
remain undetected. We can utilize this information and investigate why
certain feature pairs are missed. For instance, in the three maps we included
in our experiments, only a single junction on a small
side road has a yield sign. We are less likely to detect middle or high traffic 
density on this road.
These insights  can be used to plan test drives or as a basis for 
analyzing the relevance of specified scenarios in some real environment. 

\medskip\noindent
\textbf{Threats to Validity}

\medskip\noindent\textit{Internal Validity.}
To test our approach, we generated data with CARLA,
as this allowed us to produce a large set of test drives
using automated scripts. 
We have not tested our approach on
a set of ground truth data to check our predicates against pre-labeled data.
However, we manually inspected rendered videos
of the generated data set to match the actual driving situations we addressed with our formulas.
Combined with manually written test cases for each predicate
and all implemented logical operators,
we are confident that the predicates are capable of detecting the scenarios
they are designed for.
Even though the binding 
operator was not used pervasively in our case study, it would not have been 
possible to express a change of lanes without the operator and we expect that 
many more complex predicates pertaining to driving require the operator.
Two examples are \emph{change of heading} and \emph{giving the right of way} in an all-way stop situation.

\medskip\noindent\textit{External Validity.}
We were able to define all relevant predicates for our experiments, but this is
not fully independent of our selection of maps and the behavior of CARLA's autopilot. 
As stated in Sect.~\ref{sec:example}, we focused on analyzing urban driving scenarios,
but the available maps in the CARLA simulator also include interstate traffic.
As other works already formalize interstate traffic (see \cite{Maierhofer2020FormalizationInterstateTraffic}),
we are confident that we are also capable of analyzing new types of maps
and traffic situations using our introduced approach. Furthermore, we 
can also record human-controlled driving behavior using the CARLA
simulator and a hardware setup containing a steering wheel, pedals
and multiple screens. These recordings produce the same file format 
as our generated scenarios and can therefore easily be analyzed. 

\medskip\noindent\textit{Concept Validity.}
When analyzing more complex situations, the specification might get too large
for our approach to be practically usable. Especially, as data from the real world
can contain errors and deviations, various complex predicates and classification trees might become necessary.
Our experiments use the perfect world perception provided by CARLA,
which removes the fuzziness of sensor data.
Analyzing real-world data requires the intelligent handling of such
fuzzy sensor data streams. Predicates then need to take into consideration that the environmental 
perception, such as object tracking, might be incorrect or imprecise. 
Previous works show that current research is already addressing certain problems
in regards to environmental perceptions~\cite{VelascoHernandez2020AutonomousDrivingArchitectures}, such as sensor fusion~\cite{Fadadu2022MultiViewFusion}, or object reference generation~\cite{Philipp2021Automated3dObject}. 
We are confident that with further results and insights,
we can use our formal specifications to include fuzzy perception data.

\section{Related Work}
\label{sec:relatedWork}
Our approach is related to various existing 
works on the safety of autonomous vehicles. 

\medskip\noindent
\textbf{Formalizing traffic scenarios.} 
Previous work formalizes traffic rules using different formal logics to define
specific scenario rules. Esterle et al. formalize traffic rules for highway
situations \cite{Esterle2020FormalizingTrafficRules} by using Linear Temporal Logic (LTL). 
The same logic is also used by 
Rizaldi et al.~\cite{Rizaldi2017FormalisingMonitoringTraffic} to formalize German overtaking rules. 
Additionally, they provide verified checkers that are able to calculate the satisfaction
of a specific trace against the defined LTL formulas. Other works formalize similar traffic rules
using the Metric Temporal Logic such as interstate traffic~\cite{Maierhofer2020FormalizationInterstateTraffic} or intersections~\cite{ Maierhofer2022FormalizationIntersectionTraffic}.
Additional traffic rules regarding uncontrolled intersections are formally described
by Karimi and Duggirala~\cite{Karimi2020FormalizingTrafficRules} using Answer Set Programming.
Most of their rules specify the expected behavior of traffic participants in regards of
the right of way at unprotected intersections. 

\medskip\noindent
\textbf{Scenario-based Testing.}
In the past few years, research has started to focus on scenario-based safety assurance
(mainly testing) of autonomous vehicles \cite{Weber2019FrameworkDefinitionLogical}, exploring definition,
specification, instantiation, execution~\cite{Dosovitskiy2017CarlaOpenUrban}, and generation of scenarios for scenario-driven development~\cite{Bock2019AdvantageousUsageTextual}, for regression testing~\cite{Rocklage2017AutomatedScenarioGeneration}, and for accident scenarios~\cite{ Jenkins2018AccidentScenarioGeneration}, mining
scenarios from data, test automation~\cite{Sun2022ScenarioBasedTest}, notions of similarity between scenarios~\cite{Zhao2021LargeScaleAutonomous}, and on
finding critical test scenarios~\cite{Abdessalem2018TestingVisionBased}. Steimle et al.~\cite{Steimle2021ConsistentTaxonomyScenario} provide a
taxonomy and definitions of terms for scenario-based development and
testing. Ulbrich et al.~\cite{Ulbrich2015DefiningSubstantiatingTerms} define a scenario
to be a sequence of scenes and a scene to be a snapshot of a
vehicle’s environment, including all actors, observers,
self-representations, and relationships between them.
Klischat and Althoff~\cite{Klischat2019GeneratingCriticalTest} generate critical test scenarios
using evolutionary algorithms by minimizing the solution space of the vehicle under test.
Menzel et al.~\cite{Menzel2018ScenariosDevelopmentTest} define that scenarios can be functional,
logical, or concrete. 
A functional scenario describes the entities
of the domain and their relation at a semantic level
(different levels of abstraction are deemed possible). Logical
scenarios represent entities and relations with the help of
parameter ranges, i.e., provide an interpretation of the semantic
signature on the tempo-spatial structures that represent scenes.
Concrete scenarios, finally, are individual instances of
tempo-spatial structures with their semantic meaning. 
These notions are widely accepted in industry and academia today and provide a
framework for formulating goals and challenges~\cite{Menzel2019FunctionalLogicalScenarios, Elster2021FundamentalDesignCriteria}. 

Generating scenarios from semantic primitives and developing adequate semantic
primitives is approached by Zhang et al.~\cite{Zhang2020MultiVehicleInteraction}
and Medrano-Berumen and Akbas~\cite{MedranoBerumen2019AbstractSimulationScenario}.
The first work generates collision-free traffic scenarios by describing road shapes using
extracted traffic primitives. 
Similarly, the second work generates roadways by connecting building blocks
(i.e., geometric primitives). In contrast to our work, these works describe scenarios
using geometric shapes and calculations to build scenarios, while we describe scenarios
with logically defined higher-level predicates.

\medskip\noindent
\textbf{Analyzing Real-World Data.}
Besides scenario-based testing being widely accepted for testing autonomous
systems, evaluations on real-world data are nevertheless mandatory~\cite{Winner2018PegasusfirstStepsSafe}.
Especially, as real-world data can be used to find relevant or critical 
traffic scenarios, which can then be applied to develop scenarios for
scenario-based testing~\cite{Menzel2018ScenariosDevelopmentTest}.
Real-world data can also be utilized to help to understand
how human drivers perceive autonomous system failures in real-world
situations~\cite{Dikmen2016AutonomousDrivingReal}. 
Other applications for real-world data are: decision making~\cite{Furda2011EnablingSafeAutonomous}, 
pedestrian intention estimation~\cite{Alvarez2020AutonomousDrivingFramework} or object
detection~\cite{Li2022CodaRealWorld}.
Nevertheless, real-world data should always be accompanied by exhaustive simulation data~\cite{Langner2018EstimatingUniquenessTest},
as it can model situations that are not feasible to test in the real-world (e.g., accidents).

\medskip\noindent
\textbf{Coverage and Metrics}.
To check whether a test set is sufficient, coverage criteria are developed.
For this, Laurent et al.~\cite{Laurent2022ParameterCoverageTesting} introduce a coverage
criterion for the parameters that are utilized in the decision process of autonomous systems.

Langner et al.~\cite{Langner2018EstimatingUniquenessTest} automatically detect novel traffic scenarios using
a machine-learning approach. 
Using this, they are able to reduce a given test set to unique test scenarios. Their future work
includes labeling of scenarios to further improve the classification of novel 
scenarios.

Hauer et al.~\cite{Hauer2019DidWeTest} introduce a test ending criterion for testing automated and autonomous 
driving systems that should help arguing over the safety of autonomous vehicles.

Closest to ours are the following two works:
Amersbach and Winner~\cite{Amersbach2019DefiningRequiredFeasible} introduce a first approach for scenario coverage by calculating the
required number of concrete scenarios regarding specified parameter ranges. They argue that
for validating highly automated vehicles a specification of functional scenarios (e.g., lane-change,
following, etc.) has to be developed. 
Li et al.~\cite{Li2022ComoptCombinationOptimization} generate abstract scenarios while maximizing the coverage of \emph{k}-way combinatorial testing. Each abstract scenario can be seen as an equivalence class for which a set of concrete scenarios is generated. The categories used for generating the abstract scenarios are similar to the scenario classifiers used in this paper (e.g., weather, road type, ego-action).
  
\section{Conclusion}
\label{sec:conclusion}

We have presented a logic for expressing features of 
driving scenarios in a temporal logic and for combining 
classifiers for features into tree-based scenario classifiers
that structure the operational design domain of an 
autonomous vehicle into relevant scenario classes.
Tree-based scenario classifiers enable an analysis of 
scenario class coverage for recorded driving data. 
We have evaluated our technique in simulated urban driving
experiments. The results show that we are capable of achieving full
coverage for some scenario classifiers and can reason about
the observed features of the analyzed set of recorded test drives.

\bibliographystyle{IEEEtranPatched}
\bibliography{IEEEabrv,references}
  
\end{document}